\documentclass[final,1p,times]{elsarticle} 
\usepackage{graphicx} 
\usepackage{amssymb} 
\usepackage{amsthm} 
\usepackage{lineno} 

\usepackage{wrapfig}   

\journal{Nuclear Physics A} 
\begin{document} 

\begin{frontmatter} 


\title{Search for the critical point of strongly interacting
matter in NA49}

\author{Katarzyna Grebieszkow$^{a}$ for the NA49 collaboration}

\address[a]{Warsaw University of Technology, 
Koszykowa 75, PL-00-662 Warsaw, Poland}

\begin{abstract} 

Theoretical calculations locate the QCD critical point at energies 
accessible at the CERN SPS. Several observables were suggested to look 
for it. Here, we present the system size dependence and the energy 
dependence of event-by-event mean transverse momentum and multiplicity 
fluctuations, as well as the energy dependence of anti-baryon to baryon 
ratios.
\end{abstract} 

\end{frontmatter} 



\section{Introduction}

\begin{wrapfigure}{r}{7.cm}
\vspace{-0.5cm}
\includegraphics[scale=0.39]{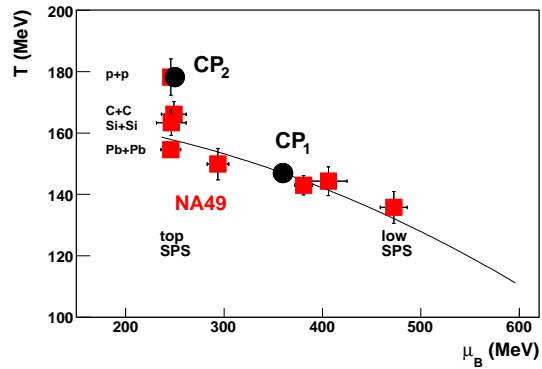}
\vspace{-0.7cm}
\caption[]{Chemical freeze-out points in the NA49 (squares) and two 
possible locations of the critical point (circles). See text for 
details.}
\label{cp1cp2}
\end{wrapfigure}

Lattice QCD calculations locate the critical point (CP) of
strongly interacting matter in the SPS energy range \cite{fodor_latt_2004}.
Among the experimental observables proposed to look for the CP are:
fluctuations of mean transverse momentum and multiplicity \cite{SRS},
transverse mass spectra of baryons and anti-baryons \cite{Askawa}, 
pion pair (sigma mode) intermittency \cite{grecy}, and elliptic flow of 
baryons and mesons~\cite{Shuryak}~\footnote 
{The energy dependence of $v_2$ for baryons and mesons was
shown in the talk. The existing data are not conclusive and better 
measurements are required for the SPS energy range.}. 
A maximum effect of the CP is expected when the freeze-out happens
near the critical point.
The $(T, \mu_B)$ phase diagram can be scanned by changing the energy 
($\mu_B$) and the size ($T$) of the colliding system (see 
Fig.~\ref{cp1cp2}). The following strategy was followed by NA49 
\cite{na49_nim} to look for the CP: an energy scan (beam energies 20$A$ - 
158$A$ GeV) with central Pb+Pb collisions, and a study of system size 
dependence (different ions) at 158$A$ GeV.

\section{Event-by-event transverse momentum and multiplicity
fluctuations}

At the CP enlarged fluctuations of multiplicity and mean transverse
momentum were suggested \cite{SRS}. In NA49 we used the $\Phi_{p_T}$
correlation measure \cite{Gaz92, fluct_size, fluct_energy} and the scaled
variance of the multiplicity distribution $\omega$ \cite{omega_size,
omega_energy} to study $p_T$ and $N$ fluctuations, respectively. 
While $\Phi_{p_T}$ is independent of $N_{part}$ fluctuations, $\omega$ 
is strongly affected by $N_{part}$ fluctuations. In order to suppress
this effect \cite{konchak} the scaled variance $\omega$ was studied for 
very central (1\%) collisions.

\pagebreak 

\begin{figure}[ht]
\centering
\vspace{-0.35cm}
\includegraphics[scale=0.6]{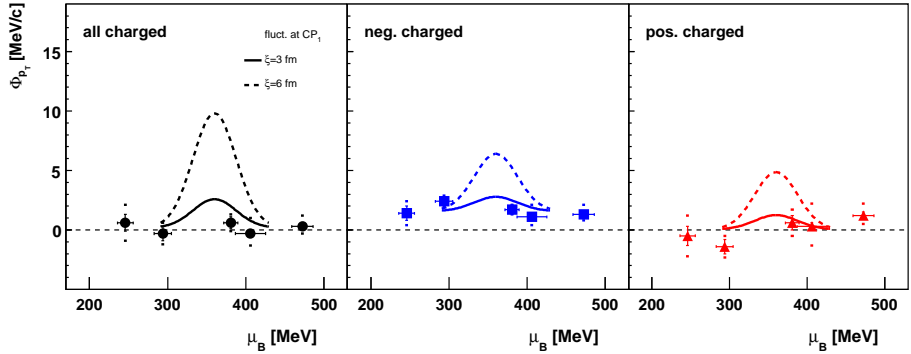}
\vspace{-0.5cm}
\caption[]{Energy dependence of $\Phi_{p_T}$ for the 7.2\% most central
Pb+Pb collisions \cite{fluct_energy} in the
forward-rapidity region $1.1 < y^{*}_{\pi} < 2.6$ and $0.005 < p_T <
1.5$ GeV/c; $y^{*}_p < y^{*}_{beam} -0.5$ (to reject the projectile
spectator domain) and common azimuthal angle acceptance. Lines correspond
to $CP_1$ predictions (see text) added to the energy averaged $\Phi_{p_T}$
measurement.}
\label{fiptmb}
\end{figure}

\begin{figure}[ht]
\centering
\vspace{-0.7cm}
\includegraphics[scale=0.6]{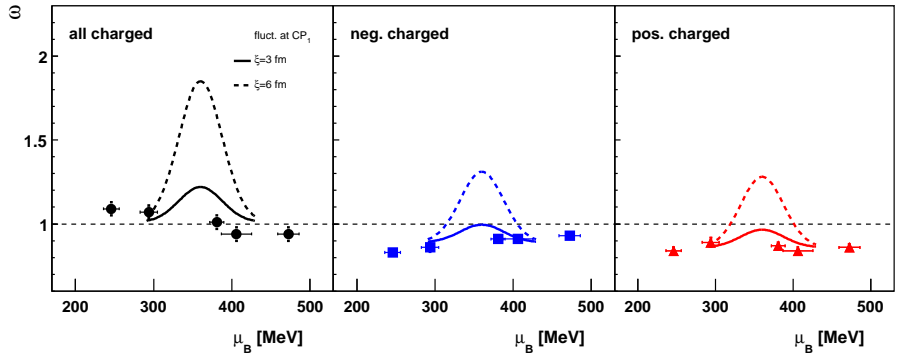}
\vspace{-0.5cm}
\caption[]{Energy dependence of $\omega$ for the 1\% most central Pb+Pb 
collisions \cite{omega_energy} in the forward-rapidity region $1.1 < 
y^{*}_{\pi} < y_{beam}$ and NA49 azimuthal angle acceptance. Lines 
correspond to $CP_1$ predictions (see text) added to the energy averaged
$\omega$ measurement.}
\label{omegamb}
\end{figure}

\begin{figure}[ht]
\centering
\vspace{-0.3cm}
\includegraphics[scale=0.6]{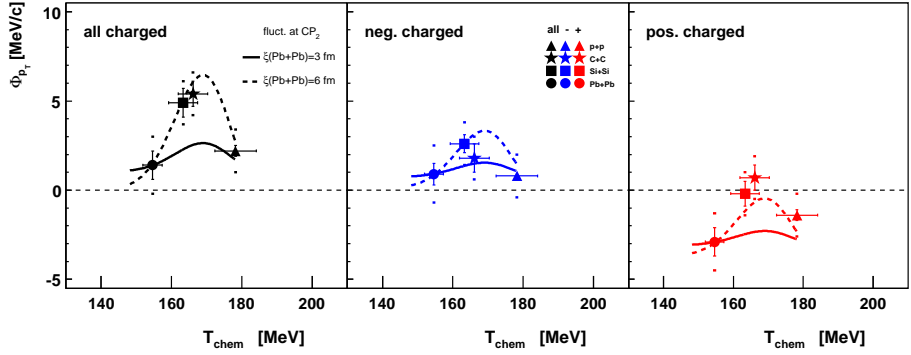}
\vspace{-0.5cm}
\caption[]{System size dependence of $\Phi_{p_T}$ at 158$A$ GeV showing results
from p+p, semi-central C+C (15.3\%) and Si+Si (12.2\%), and 5\% most 
central Pb+Pb collisions \cite{fluct_size}.
Forward-rapidity region $1.1 < y^{*}_{\pi} < 2.6$ and $0.005 < p_T <
1.5$ GeV/c; NA49 azimuthal angle acceptance. Lines correspond to
$CP_2$ predictions (see text) shifted to reproduce the $\Phi_{p_T}$ value
for central Pb+Pb collisions.}
\label{fiptT}
\end{figure}

\begin{figure}[ht]
\centering
\vspace{-0.3cm}
\includegraphics[scale=0.6]{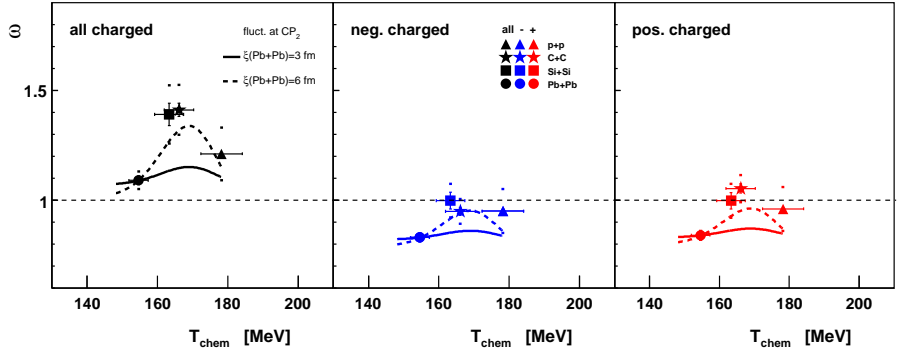}
\vspace{-0.5cm}
\caption[]{System size dependence of $\omega$ at 158$A$ GeV for the 1\% most 
central p+p \cite{omega_size}, C+C and Si+Si \cite{benjaminPhD}, and 
Pb+Pb collisions \cite{omega_energy}.
Forward-rapidity region $1.1 < y^{*}_{\pi} < y_{beam}$ ($1.1 <
y^{*}_{\pi} < 2.6$ for p+p); NA49 azimuthal angle acceptance. Lines
correspond to $CP_2$ predictions (see text) shifted to reproduce the 
$\omega$ value for central Pb+Pb collisions. }
\label{omegaT}
\end{figure}

Figures \ref{fiptmb}, \ref{omegamb}, \ref{fiptT}, and \ref{omegaT} 
present the energy ($\mu_B$) and system size
($T_{chem}$) dependence of $\Phi_{p_T}$ and $\omega$. The chemical
freeze-out parameters, $T_{chem}(A,\sqrt{s_{NN}})$ and
$\mu_B(A,\sqrt{s_{NN}})$ were taken from fits of the hadron gas model 
\cite{beccatini} to particle yields. The lines correspond to critical 
point predictions with the magnitude of fluctuations at the CP taken 
from Ref.\cite{SRS, MS} assuming correlation lengths $\xi$ decreasing 
monotonically
with decreasing system size: a) $\xi$(Pb+Pb) = 6 fm and  $\xi$(p+p) = 2 
fm (dashed lines) or b) $\xi$(Pb+Pb) = 3 fm and  $\xi$(p+p) = 1 fm 
(solid lines). The predictions include corrections by NA49 due
to the limited rapidity range (forward-rapidity) and azimuthal
angle acceptance of the detector. The width of the enhancement due to 
the CP in the ($T, \mu_B$) plane
is based on Ref. \cite{hatta} and taken as $\sigma (\mu_B) \approx 30$
MeV and $\sigma (T) \approx 10$ MeV. 
We considered two possible locations of the critical point as shown in
Fig.~\ref{cp1cp2}: $\mu_B (CP_1)$ taken from lattice QCD calculations  
\cite{fodor_latt_2004} and $CP_2$ assuming 
that the chemical freeze-out point of p+p data at 158$A$ GeV 
may be located on the phase transition line.

Figures \ref{fiptmb} and \ref{omegamb} show no significant peak in the 
energy dependence of $\Phi_{p_T}$ and $\omega$ at SPS energies. These 
data therefore show no indication for critical point fluctuations, 
however a narrower $\mu_B$ scan would be desirable. Figures \ref{fiptT} 
and \ref{omegaT} display a maximum of $\Phi_{p_T}$ and $\omega$ for C+C 
and Si+Si interactions at the top SPS energy. This peak is two times
higher for all charged than for negatively charged particles, as
expected for the effect of the CP \cite{SRS}. Both figures suggest 
that the NA49 data are consistent with the predictions for $CP_2$ 
(curves).

\section{Transverse mass spectra of baryons and anti-baryons}

The presence of a CP can deform the evolution trajectories
of the expanding fireball in the $(T, \mu_B)$ phase
diagram (focusing effect) \cite{Askawa}. This is expected
to result in a decrease of
the ${ \bar{p}} / {p}$ ratio (generally anti-baryon to baryon ratio
${\bar{B}} / {B}$) with increasing transverse
momentum instead of a rise or flat behavior in a scenario without the 
critical point.

\begin{figure}[ht]
\centering
\vspace{-0.4cm}
\includegraphics[width=0.4\textwidth]{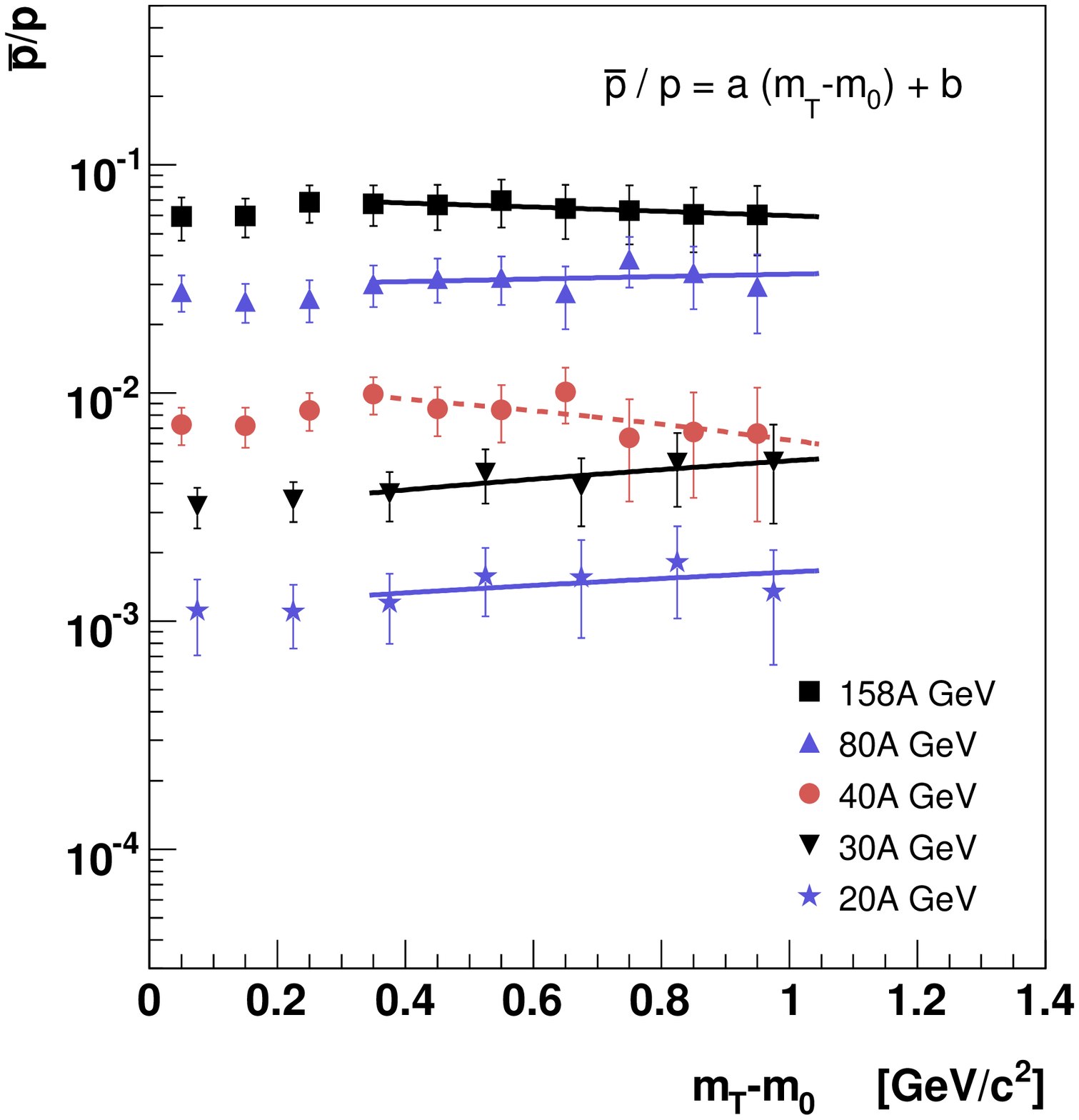}
\includegraphics[width=0.45\textwidth]{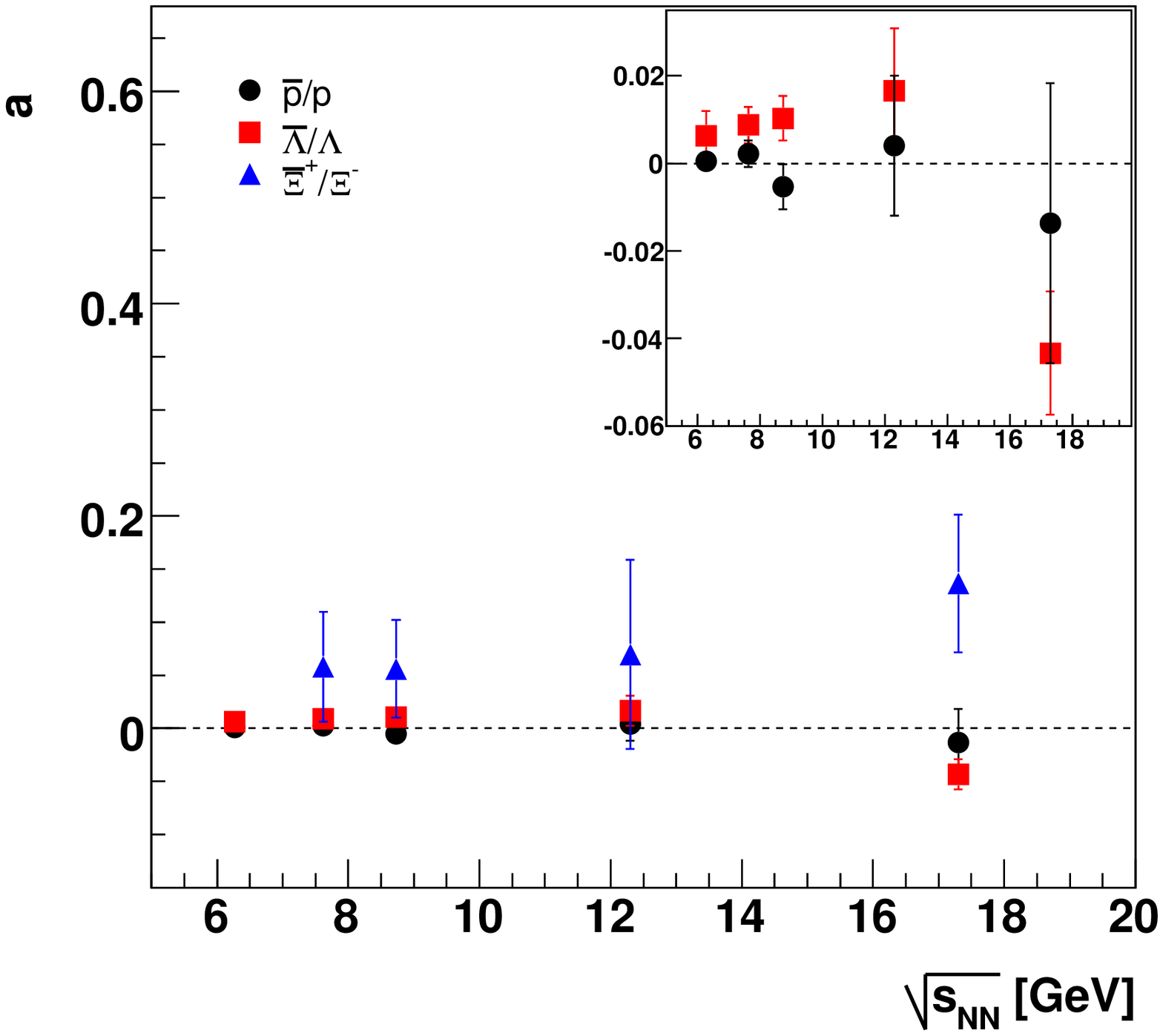}
\vspace{-0.5cm}
\caption[]{Left: ${ \bar{p}} / {p}$ ratio versus
reduced transverse mass $m_T-m_0$. Right: slope parameters $a$ of the
${ \bar{B}} / {B}$ ratios. The inset has an expanded vertical scale.}
\label{mtspectra_a}
\end{figure}

Figure~\ref{mtspectra_a}~(left) presents the ${ \bar{p}} / {p}$ ratio 
versus reduced transverse mass $m_T-m_0$. The NA49 measurements were 
taken from Ref. \cite{mtspect_p}; the plotted data points were obtained 
by dividing the $\bar{p}$ data by values from curves fitted to the 
proton results. The lines show linear fits to ${ \bar{p}} / {p}$
versus $m_T-m_0$  with $a$ being the slope parameter. The
dashed line for 40$A$ GeV (with slightly higher slope) was originally
\cite{Askawa} considered the best indication for the predicted critical
effect. A similar analysis was also done for
${ \bar{\Lambda}} / {\Lambda}$ and ${ {\bar{\Xi}}^{+}} / {\Xi^{-}}$
\cite{mtspect_l}. The slope parameters $a$ of all three ${\bar{B}} / 
{B}$ ratios are presented in Fig.~\ref{mtspectra_a}~(right) 
and show no significant energy dependence. This observation for 
the ${ \bar{p}} / {p}$ ratio remains valid also when the energy
range is extended to the RHIC domain~\cite{bedanga}. Thus we conclude
that transverse mass spectra of $B$ and ${\bar{B}}$ show no evidence
for the critical point.

\section{Summary and Conclusions}

NA49 results show no
indication of the critical point in the energy dependence of
multiplicity and mean transverse momentum fluctuations, and ratios of 
the anti-baryon/baryon transverse mass spectra in central Pb+Pb 
collisions.
A maximum of mean $p_T$ and multiplicity fluctuations as a function of
the system size is observed at 158$A$ GeV for smaller systems (Si+Si,
C+C). If one takes the results as an indication of the CP it would be
located at $T \approx 178$ MeV and $\mu_B \approx 250$ MeV. However, a
detailed energy and system-size scan is necessary to establish the
existence of the critical point.




\end{document}